\documentclass[preprints,article,accept,oneauthor,pdftex]{Definitions/mdpi} 

\firstpage{1} 
\pubvolume{}
\issuenum{}
\articlenumber{}
\pubyear{}
\copyrightyear{}
\history{} 

\pdfoutput=1

\usepackage{braket}
\DeclareMathOperator{\Tr}{Tr} 

\renewcommand*{\Re}{\mathop{\mathrm{Re}}\nolimits}

\Title{Long-Term Behaviour in an Exactly Solvable Model of Pure Decoherence and the Problem of Markovian Embedding}

\Author{Anton Trushechkin $^{1,2,3,\dag}$}

\AuthorNames{Anton Trushechkin}

\address{%
$^{1}$ \quad Steklov Mathematical Institute of Russian Academy of Sciences, Department of Mathematical Physics, Gubkina 8, Moscow 119991, Russia,\\
$^{2}$ \quad Heinrich Heine University D\"{u}sseldorf,
Faculty of Mathematics and Natural Sciences, Institute for Theoretical Physics III, 
Universit\"{a}tsstr. 1, D\"{u}sseldorf 40225, Germany,\\
$^{3}$ \quad National University of Science and Technology MISIS, Department of Mathematics, Leninskiy Prospekt 4, Moscow 119049, Russia,\\
$^\dag$ \quad trushechkin@mi-ras.ru}

\corres{Correspondence: trushechkin@mi-ras.ru}

\abstract{We consider a well-known exactly solvable model of an open quantum system with pure decoherence. The aim of this paper is twofold. Firstly,  decoherence is a property of open quantum systems important for both quantum technologies and the fundamental question of quantum-classical transition. It is worthwhile to study how the long-term rate of decoherence depends on the spectral density characterizing the system-bath interaction in this exactly solvable model. Secondly, we address a more general problem of the Markovian embedding of a non-Markovian open system dynamics. It is often assumed that a non-Markovian open quantum system can be embedded into a larger Markovian system. However, we show that such embedding is possible only for the Ohmic spectral densities (for the case of a positive bath temperature) and is impossible for both the sub- and super-Ohmic spectral densities. From the other side, for the Ohmic spectral densities, an asymptotic large-time Markovianity (in terms of the quantum regression formula) takes place.
}

\keyword{open quantum systems, decoherence, Markovian and non-Markovian dynamics} 

\PACS{03.65.Yz}
\MSC{81S22}
\begin{document}


\section{Introduction}

Decoherence is a fundamental property of quantum systems subject to external noise. It is important from both fundamental and practical point of view. Fundamentally, decoherence is often considered as the phenomenon responsible for the emergence of the description of macroscopic physical reality based on classical mechanics and classical (Kolmogorov) probability theory from the quantum description. Quantum coherence is a distinctive property of the quantum world and the decay of coherence is at least one of the factors of quantum-to-classical transition. From the viewpoint of applications, decoherence is one of the obstacles for construction of a quantum computer, which requires preservation of coherence during the computation.

Mathematically, quantum system subject to external noise are studied within the framework of the theory of open quantum systems, in which models of systems interacting with a thermal bath or another environment are considered. In this paper, we will study a particular well-known exactly-solvable model of pure decoherence \cite{BP,EkertQCompDiss,AlickiDecoh}. Despite of the fact that it is well-known and attract interests in a number of recent studies (especially with respect to the question of Markovianity or non-Markovianity of the dynamics) \cite{Brito,VacchiniQRT,MerkliNesterovDimer,LonigroChrus,SuperOhmic},  precise asymptotic estimations of the decoherence rate depending on the spectral density (specifying the system-bath interaction) are absent. The purpose of this paper is to fill this gap.

Then we discuss some consequences of these results, which can be important for the present active discussions about non-Markovian open quantum dynamics. There is a hierarchy of mathematical definitions related to our intuitive understanding of ``Markovianity'' or the ``absence of memory of the bath'' in the context of open quantum systems \cite{MarkHier}. According to one approach, Markovianity is associated with the semigroup property of the dynamics. A generator of this semigroup has the well-known form of Gorini-Kossakowski-Sudarshan-Lindblad (GKSL; we consider only finite-dimensional systems). So, in this case, the dynamics of the reduced density operator of the system is rather simple and can be described by a GKSL quantum master equation, which, in the finite-dimensional case, is simply a system of linear ordinary differential equations with constant coefficients. Definitions of Markovianity related to certain properties of the dynamics of the reduced density operator of the system, like decrease of the distinguishability of the evolving quantum states or CP-divisibility (``CP'' means ``completely positive'') cover wider ranges of models and include certain classes of time-dependent GKSL generators \cite{MarkHier,VacchiniQRT,ChrusIntroNonMark,ChrusBeyondMark}. The most general (and the strongest) definition of Markovianity, which generalizes the corresponding definition from the classical random process theory, is related to the quantum regression formula \cite{MarkHier,VacchiniQRT,ChrusBeyondMark,Lax,LoGullo}.

\begin{figure}[H]
\begin{center}
\includegraphics[width=12 cm]{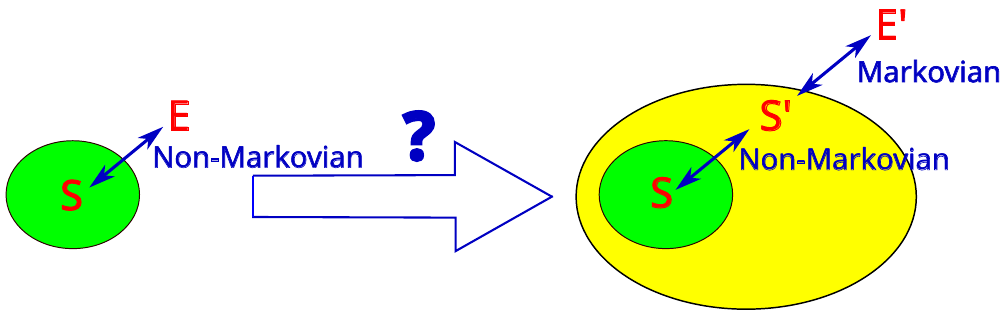}
\end{center}
\caption{Schematic representation of the assumption of Markovian embedding. An interaction of an open quantum system ${\rm S}$ with an environment ${\rm E}$ gives rise to non-Markovian dynamics of the system. Is there an extension ${\rm S'}$ of the system (either physical or fictitious) such that the dynamics of the enlarged system ${\rm S+S'}$ (interacting with the residue environment ${\rm E'}$) is Markovian? \label{fig1}}
\end{figure}

One of the popular ways of dealing with non-Markovian quantum dynamics is to embed an open quantum system to a larger system whose dynamics is Markovian in the sense that its dynamics is described by a quantum dynamical semigroup with a (time-independent) GKSL generator. Of course, the original system-bath unitary dynamics is obviously Markovian. But the bath typically contains a continuum number of oscillatory modes. In the described approach, we try to embed our system into a larger system, which is still finite-dimensional or, at least, has a discrete energy spectrum, so that it is possible to cut high energy levels and approximately reduce its (Markovian) dynamics to a finite-dimensional (also Markovian) dynamics. The methods of pseudomodes \cite{Tamapre,Tama,GarrawayPetruc,TereFinT,TereSeveralBath} and reaction (collective) coordinate \cite{Lambert,Strasberg,Segal} are examples of this approach. The assumption of a possibility of a Markovian embedding is often used in data-driven prediction of the dynamics of open quantum systems \cite{Luchnikov19,Luchnikov22}. The method of Markovian embedding is schematically depicted on Fig.~\ref{fig1}. However, we show that, for some bath spectral densities, such embedding is impossible. Namely, in the case of a positive temperature of the bath, only for the case of an Ohmic spectral density (i.e., very specific asymptotic behaviour in a vicinity of zero), the Markovian embedding is possible. But even in this case, there can be factors not described by Markovian embeddings. If the spectral density is either sub- or super-Ohmic (which are common in physics), such embedding is impossible.

Vice versa, for the Ohmic spectral density (again in the case of a positive temperature), we can observe the asymptotic Markovianity in the strongest sense related to the quantum regression formula. That is, even if the dynamics is non-Markovian, it becomes Markovian on large times, which simplifies its description. This results can be considered as a development of the results by D.~Lonigro and D.~Chru\'{s}ci\'{n}ski \cite{LonigroChrus}. Namely, they show that the quantum regression formula for the considered pure decoherence model is  satisfied exactly if the bath spectral density is flat for the whole real line of positive and negative frequencies of the bath. However, these conditions (especially negative frequencies) are unphysical. The authors, however, note that the flat spectral density can be a reasonable approximation in realistic scenarios. Here we confirm this hypothesis: asymptotic satisfiability of the quantum regression formula is explained by the fact that, for large times, only the behaviour of the spectral density in a vicinity of zero matters, so, it can be considered to be approximately flat on the whole real line.

The following text is organized as follows. In Sec.~\ref{SecProblem}, we describe the model. In Sec.~\ref{SecMain}, we prove the main theorem about the long-term rates of decoherence depending on the asymptotic behaviour of the spectral density in a vicinity of zero. In Sec.~\ref{SecExamples}, we illustrate the theorem by two particular widely used spectral densities. In Sec.~\ref{SecWeak}, we compare the obtained results with various rigorous results about the Davies GKSL generator of a quantum dynamical semigroup in the weak-coupling regime \cite{Davies,Davies2,MerkliRev}. Sec.~\ref{SecMarkEmb} is devoted to the impossibility of the Markovian embedding in the cases of sub- or super-Ohmic spectral densities. In Sec.~\ref{SecAsympMark}, we prove asymptotic Markovianity (in the sense of the quantum regression formula) for the Ohmic spectral densities. In all these sections we consider the realistic case of a positive temperature of the bath. However, the zero-temperature bath is often used as a reasonable approximation when the temperature is low. The results can be easily transferred to this case as well, which is discussed in Sec.~\ref{SecZero}.

\section{Problem statement}\label{SecProblem}

Let us consider a two-dimensional quantum system (a qubit) interacting with a finite number of $N$ quantum harmonic oscillators. Mathematically speaking, we work in the Hilbert space $\mathcal H_{\rm S}\otimes\mathcal H_{\rm B}=\mathbb C^2\otimes(\ell^2)^{\otimes N}$. Consider the following Hamiltonian (self-adjoint operator) acting in this space:
\begin{equation}\label{EqH}
H=H_{\rm S}+H_{\rm B}+H_{\rm I}=
\frac{\omega_0}2\sigma_z+\sum_{k=1}^N\omega_ka_k^\dag a_k+
\frac{\sigma_z}2\otimes\sum_{k=1}^N (g_ka_k^\dag+g_k^*a_k),
\end{equation}
where 
\begin{equation}
\sigma_z=
\begin{pmatrix}
1&0\\0&-1
\end{pmatrix}
=
\ket1\bra1-\ket0\bra0,
\quad
\ket1=\begin{pmatrix}
1\\0
\end{pmatrix},
\quad
\ket0=\begin{pmatrix}
0\\1
\end{pmatrix}
\end{equation}
are one of the Pauli matrices and basis elements (the standard basis) of $\mathbb C^2$, respectively. Further, $a_k^\dag$ and $a_k$ are the bosonic creation and annihilation operators, $\omega_k>0$, $\omega_0$ is a real number and $g_k$ are complex numbers.

Consider the following initial density operator (quantum state):
\begin{equation}\label{EqRhoIni}
\rho(0)=\rho_{\rm S}(0)\otimes\rho_{{\rm B},\beta}\equiv
\rho_{\rm S}(0)\otimes Z_{\rm B}^{-1}e^{-\beta H_{\rm B}},
\end{equation}
where $\rho_{\rm S}(0)$ is the initial density operator of the system and $\rho_{{\rm B},\beta}$ is the thermal (Gibbs) state of the bath with $\beta$ being the inverse temperature, $Z_{\rm B}=\Tr e^{-\beta H_{\rm B}}$. The joint system-bath state at time $t$ is given by
\begin{equation}
\rho(t)=e^{-itH}\rho(0)e^{itH}.
\end{equation}
We are interested in the reduced density operator of the system:
\begin{equation}
\rho_{\rm S}(t)=\Tr_{\rm B}\rho(t)=\Tr_{\rm B}[e^{-itH}\rho(0)e^{itH}],
\end{equation}
where $\Tr_{\rm B}$ denotes the partial trace over the space $\mathcal H_{\rm B}$. The interaction representation:
\begin{equation}\label{EqVarrho}
\varrho(t)=
e^{it H_{\rm S}}\rho_{\rm S}(t)e^{-itH_{\rm S}}=
\begin{pmatrix}
\varrho_{11}(t)&\varrho_{10}(t)\\
\varrho_{01}(t)&\varrho_{00}(t)
\end{pmatrix},
\end{equation}
where $\varrho_{jk}$ are the matrix elements of the operator $\varrho$ in the standard basis $\{\ket1,\ket0\}$.

Since $[H,\sigma_z]=0$, $\varrho_{11}(t)=\varrho_{11}(0)$ and $\varrho_{00}(t)=\varrho_{00}(0)$. Also, 
\begin{equation*}
\varrho_{10}(t)=
\varrho_{10}(0)
\Tr_{\rm B}
\big[e^{-itH_1}\rho_{{\rm B},\beta}e^{itH_0}\big],
\end{equation*}
where
\begin{equation}\label{EqH10}
H_j=\sum_{k=1}^N\omega_ka_k^\dag a_k+
\frac{(-1)^{j-1}}2
\sum_{k=1}^N (g_ka_k^\dag+g_k^*a_k),
\end{equation}
$j=0,1$.
One can show \cite{BP,AlickiDecoh,EkertQCompDiss,LonigroChrus,ChrusBeyondMark} that
\begin{equation}\label{EqExpGamma}
\Tr_{\rm B}
\big[e^{-itH_1}\rho_{{\rm B},\beta}e^{itH_0}\big]
=
e^{-\Gamma(t)},
\end{equation}
where
\begin{equation}\label{EqGamma}
\begin{split}
\Gamma(t)
&=
\sum_{k=1}^N|g_k|^2\coth\left(\frac{\beta\omega_k}2\right)
\frac{1-\cos\omega_k t}{\omega_k^2}
\\
&=\int_0^\infty J(\omega)\coth\left(\frac{\beta\omega}2\right)
\frac{1-\cos\omega t}{\omega^2}d\omega
\\&=
\int_0^\infty J_{\rm eff}(\omega)
\frac{1-\cos\omega t}{\omega^2}d\omega.
\end{split}
\end{equation}
Here 
\begin{equation}\label{EqJ}
J(\omega)=\sum_{k=1}^N |g_k|^2\delta(\omega-\omega_k)
\end{equation}
is the spectral density function of the bath and 
\begin{equation}
J_{\rm eff}(\omega)=
J(\omega)\coth\left(\frac{\beta\omega}2\right)
\end{equation}
is sometimes called the effective spectral density function.

The diagonal elements of the density matrix $\varrho_{11}(t)$ and $\varrho_{00}(t)$ are called the \textit{populations} and the off-diagonal elements $\varrho_{10}(t)$ and $\varrho_{01}(t)=\varrho_{10}(t)^*$ are called the \textit{coherences}. The decrease of the off-diagonal elements is called \textit{decoherence}. It is clear that $\Gamma(t)\geq\Gamma(0)=0$ and, hence, $|\varrho_{10}(t)|\leq|\varrho_{10}(0)|$. If there finitely many oscillators $N$ in the bath, then $\Gamma(t)$ and $\varrho(t)$ are periodic or almost periodic functions of $t$.

In the thermodynamic limit of the bath $N\to\infty$, the index
$k$ runs over a continuous set and we assume that $J(\omega)$ tends to an integrable function on the half-line $[0,\infty)$. Note that there is a mathematically rigorous way to start directly from the continuum of oscillators rather than to perform a thermodynamic limit of a finite number of oscillators \cite{MerkliIdealQGas,TMCA}. However, for our purposes, this mathematically simpler approach with the thermodynamic limit will suffice.

We will be interested in the behaviour of $\Gamma(t)$ given by the second (or third) line of Eq.~(\ref{EqGamma}), where $J(\omega)$ is an integrable function on $[0,\infty)$, for large times $t\to\infty$. If $\Gamma(t)\to\infty$ and, thus, $\varrho_{10}(t)\to0$ as $t\to\infty$, then we will speak that the \textit{full decoherence} occurs. If $\Gamma(t)$ is a bounded function of $t$, so, $|\varrho_{10}(t)|\geq\varrho_{10}^*>0$, then we will speak that the \textit{partial decoherence} occurs. The partial decoherence for certain classes of the spectral densities is a known theoretical prediction \cite{EkertQCompDiss,AlickiDecoh,Viola2013,SuperOhmic}. For completeness, we include this results into Theorem~\ref{ThMain}, but mainly we are interested in the case of the full decoherence. Then the question we are interested in is the following: \textit{What is the asymptotic rate of convergence of $\varrho_{10}(t)$ to zero for large $t$ depending of the properties of $J(\omega)$}? In book \cite{BP} and also in Ref.~\cite{EkertQCompDiss}, only a particular forms of $J(\omega)$ are considered. Here we obtain a general answer.

The crucial is the asymptotic behaviour of $J(\omega)$ for small $\omega$: Pure decoherence is caused by interaction with the low frequencies of the bath. We consider only the case $J(\omega)\sim c\omega^\gamma$ as $\omega\to+0$ (i.e. $\lim_{\omega\to+0}J(\omega)\omega^{-\gamma}=c$), where $c,\gamma>0$. If $0<\gamma<1$, $\gamma=1$ or $\gamma>1$, the spectral density is called sub-Ohmic, Ohmic or super-Ohmic, respectively.

\section{Main theorem}\label{SecMain}

\begin{Theorem}\label{ThMain}

Let $J(\omega)$ be an integrable function on $[0,\infty)$.

\begin{enumerate}
\item If $J(\omega)\sim c\omega^{2+\delta}$ as $\omega\to+0$, where $\delta>0$ and $c>0$, then
\begin{equation}\label{EqGammaConst}
\lim_{t\to\infty}\Gamma(t)=
\int_0^\infty \frac{J(\omega)}{\omega^2}\coth\left(\frac{\beta\omega}2\right)d\omega\equiv\Gamma_{\infty},
\end{equation}
hence, the decoherence is partial:
\begin{equation}\label{EqDecohPartial}
\varrho_{10}(t)=\varrho_{10}(0)B(t)e^{-\Gamma_\infty},
\end{equation}
where $B(t)$ is a bounded function converging to 1 as $t\to\infty$. 

\item If $J(\omega)\sim c\omega$ as $\omega\to+0$, where $c>0$, and there exits $J'_{\rm eff}(0)=\lim_{\omega\to+0}J'_{\rm eff}(\omega)$, then
\begin{equation}\label{EqGammaLin}
\Gamma(t)=\Gamma_0t+\alpha\ln t+C+o(1),\quad t\to\infty,
\end{equation}
where $C$ is a constant, hence, the decoherence is exponential:
\begin{equation}\label{EqDecohExp}
\varrho_{10}(t)=\varrho_{10}(0)B(t) e^{-\Gamma_0 t}/t^\alpha.
\end{equation}
Here
\begin{equation}\label{EqGamma0}
\begin{split}
\Gamma_0&=\frac\pi2 J_{\rm eff}(0)=
\pi\beta^{-1}\lim_{\omega\to+0}\frac{J(\omega)}\omega,
\\
\alpha&=J'_{\rm eff}(0)=2\beta^{-1}
\lim_{\omega\to+0}\left(\frac{J(\omega)}\omega\right)'
\end{split}
\end{equation}
and $B(t)$ is a bounded function converging to $e^{-C}$ as $t\to\infty$.

\item If $J(\omega)\sim c\omega^2$ as $\omega\to+0$, where $c>0$, and there exits $J'_{\rm eff}(0)=\lim_{\omega\to+0}J'_{\rm eff}(\omega)$, then the decoherence obeys a power law:
\begin{equation}\label{EqGammaLog}
\Gamma(t)=\alpha\ln t+C+o(1),\quad t\to\infty,
\end{equation}
hence, 
\begin{equation}\label{EqDecohPower}
\varrho_{10}(t)=\varrho_{10}(0)B(t)/t^\alpha,
\end{equation}
where, again, $B(t)$ is a bounded function converging to $e^{-C}$ as $t\to\infty$.

\item If $J(\omega)=\omega^{1+\delta}G(\omega)$ as $\omega\to+0$, where $-1<\delta<1$, $G(0)>0$, and there exits a derivative of the function $\omega^{-\delta}J_{\rm eff}(\omega)$ at $\omega=0$ (which, as before, can be defined in terms of the limit $\omega\to+0$), then
\begin{equation}\label{EqGammaSubLin}
\Gamma(t)=At^{1-\delta}+\widetilde C+o(1),\quad t\to\infty,
\end{equation}
if $0<\delta<1$, and
\begin{equation}\label{EqGammaSuperLin}
\Gamma(t)=At^{1-\delta}+
2\beta^{-1}G'(0)
O(t^{-\delta}\ln t),\quad t\to\infty,
\end{equation}
if $-1<\delta\leq0$,
where
\begin{equation*}
A=
2\beta^{-1}G(0)
\int_0^{\infty}
\frac{1-\cos\upsilon}{\upsilon^{2-\delta}}\,d\upsilon
\end{equation*}
and $\widetilde C$ is a constant.
Thus, the decoherence is subexponential for $\delta>0$ and superexponential for $\delta<0$:
\begin{equation}\label{EqDecohSubExp}
\varrho_{10}(t)=\varrho_{10}(0)\widetilde B(t)e^{-At^{1-\delta}}
\end{equation}
if $0<\delta<1$, where $\widetilde B(t)$ is a bounded function converging to $e^{-\widetilde C}$ as $t\to\infty$, and
\begin{equation}\label{EqDecohSuperExp}
\varrho_{10}(t)=\varrho_{10}(0)
e^{-At^{1-\delta}+2\beta^{-1}G'(0) O(t^{-\delta}\ln t)},
\quad t\to\infty
\end{equation}
if $-1<\delta\leq0$.
\end{enumerate}
\end{Theorem}

The results of Theorem~\ref{ThMain} can be summarized as follows. Let $J(\omega)\sim c\omega^\gamma$ (or, equivalently, $J_{\rm eff}(\omega)\sim c\omega^{\gamma-1}$) as $\omega\to+0$ for some $c>0$.
\begin{itemize}
\item If $0<\gamma<1$ (sub-Ohmic spectral density), then the decoherence is full and its rate is superexponential.

\item If $\gamma=1$ (Ohmic spectral density), then the decoherence is full and its rate is exponential.

\item If $1<\gamma<2$ (super-Ohmic spectral density), then the decoherence is full and its rate is subexponential, but faster than any degree of $t$.

\item If $\gamma=2$ (super-Ohmic spectral density), then the decoherence is full and obeys a power law.

\item If $\gamma>2$ (super-Ohmic spectral density), then the decoherence is partial. 
\end{itemize}

Also it can be noticed that all the decoherence constants in Theorem~\ref{ThMain} are proportional to the temperature $\beta^{-1}$, which corresponds to the physical intuition that the environment with higher temperature causes faster decoherence.

\begin{proof}[Proof of Theorem~\ref{ThMain}]
Let us consider all the cases.

\textit{Case 1.} In this case, $J_{\rm eff}(\omega)\sim\omega^{1+\delta}$, $\delta>0$, and $J_{\rm eff}(\omega)/\omega^2\sim\omega^{-(1-\delta)}$. That is, $\frac{J_{\rm eff}(\omega)}{\omega^2}$ is an integrable function. Then, by the Riemann-Lebesgue lemma, $\int_0^\infty \frac{J_{\rm eff}(\omega)}{\omega^2}\cos\omega t\,d\omega$ tends to zero as $t\to\infty$, from which Eq.~(\ref{EqGammaConst}) follows. 

Note that if we impose additional conditions on the degree of smoothness of $J_{\rm eff}(\omega)$ to provide the possibility of a certain number of iterative integrations by parts, then the long-term rate of convergence of $\Gamma(t)$ to $\Gamma_\infty$ can be estimated.

\textit{Cases 2 and 3}.  In this cases, we can express
\begin{equation}\label{EqMainDecomposition}
\begin{split}
\Gamma(t)
&=\int_0^{\omega_{\rm c}}
\frac{J_{\rm eff}(\omega)-J_{\rm eff}(0)-J'_{\rm eff}(0)\omega}{\omega^2}
(1-\cos\omega t)\,d\omega\\
&+J_{\rm eff}(0)\int_0^{\omega_{\rm c}}
\frac{1-\cos\omega t}{\omega^2}\,d\omega\\
&+J'_{\rm eff}(0)\int_0^{\omega_{\rm c}}
\frac{1-\cos\omega t}{\omega}\,d\omega\\
&+\int_{\omega_{\rm c}}^\infty
J_{\rm eff}(\omega)\frac{1-\cos\omega t}{\omega^2}\,d\omega,
\end{split}
\end{equation}
where $\omega_{\rm c}>0$ (``c'' from ``cutoff'') is arbitrary.
Consider all the terms. The first factor in the integrand of the first term is a locally integrable function and, hence, by the Riemann-Lebesgue lemma, the first term tends to a constant. The same is true for the last term: $J_{\rm eff}(\omega)/\omega^2$ is integrable on Borel sets that do not include zero. Consider the second term:
\begin{equation*}
\begin{split}
\int_0^{\omega_{\rm c}}
\frac{1-\cos\omega t}{\omega^2}d\omega
&=
\int_0^\infty
\frac{1-\cos\omega t}{\omega^2}d\omega
-\int_{\omega_{\rm c}}^\infty
\frac{1-\cos\omega t}{\omega^2}d\omega
\\&=
t\int_0^{\infty}
\frac{1-\cos\upsilon}{\upsilon^2}\,d\upsilon
-\frac1{\omega_{\rm c}}
+\int_{\omega_{\rm c}}^\infty
\frac{\cos\omega t}{\omega^2}d\omega
\\&=
\frac\pi2t
-\frac1{\omega_{\rm c}}
+\int_{\omega_{\rm c}}^\infty
\frac{\cos\omega t}{\omega^2}d\omega.
\end{split}
\end{equation*}
Here, in the second equality, we have performed the substitution $\omega t=\upsilon$. The last integral converges to zero as $t\to\infty$ by the Riemann-Lebesgue lemma. 

In the second term of (\ref{EqMainDecomposition}), we perform the same substitution $\omega t=\upsilon$ and then use a known formula  for the integral of $(1-\cos\upsilon)/\upsilon$ (see, e.g., Ref.~\cite{GR}):
\begin{equation}
\int_0^{\omega_{\rm c}}
\frac{1-\cos\omega t}{\omega}\,d\omega
=
\int_0^{\omega_{\rm c}t}
\frac{1-\cos\upsilon}{\upsilon}\,d\upsilon=
\ln\omega_{\rm c}t+\gamma-{\rm Ci}(t),
\end{equation}
where $\gamma$ is the Euler-Mascheroni constant and 
\begin{equation}
{\rm Ci}(t)=-\int_t^\infty\frac{\cos x}x\,dx
\end{equation}
is the cosine integral, which is a bounded function converging to zero as $t\to\infty$.

The expressions for the constants $\Gamma_0$ and $\alpha$ follow from the previous calculations and the following observations:
\begin{equation*}
\begin{split}
J_{\rm eff}(0)&=
\lim_{\omega\to+0}J(\omega)\coth\frac{\beta\omega}2
=
2\beta^{-1}\lim_{\omega\to+0}\frac{J(\omega)}\omega,
\\
J'_{\rm eff}(0)&=
\lim_{\omega\to+0}
\left(J(\omega)\coth\frac{\beta\omega}2\right)'
=
2\beta^{-1}\lim_{\omega\to+0}
\left(
\frac{J(\omega)}\omega
\right)'.
\end{split}
\end{equation*}
We have proved the cases~2 and~3 of the theorem. 

Note that, if $J'_{\rm eff}(0)=0$ (which is the case, e.g., for the Drude-Lorentz spectral density, see below), then we do not need to introduce a cutoff frequency $\omega_{\rm c}$ since both the first and the second integrals in Eq.~(\ref{EqMainDecomposition}) are convergent even for an infinite upper limit of integration in this case.

\textit{Case 4}. In this case, $J'_{\rm eff}(0)$ is infinite unless $\delta=0$ (i.e., case~2). However, define $\widetilde G(\omega)=\omega \coth(\beta\omega/2)G(\omega)$, so that
$J_{\rm eff}(\omega)=\omega^{\delta}\widetilde G(\omega)$. We can apply an expansion analogous to Eq.~(\ref{EqMainDecomposition}) if we apply  Taylor's formula for $\widetilde G$:
\begin{equation}\label{EqMainDecompositionG}
\begin{split}
\Gamma(t)
&=\int_0^{\omega_{\rm c}}
\frac{\widetilde G(\omega)-\widetilde G(0)-\widetilde G'(0)\omega}{\omega^{2-\delta}}
(1-\cos\omega t)\,d\omega\\
&+\widetilde G(0)\int_0^{\omega_{\rm c}}
\frac{1-\cos\omega t}{\omega^{2-\delta}}\,d\omega\\
&+\widetilde G'(0)\int_0^{\omega_{\rm c}}
\frac{1-\cos\omega t}{\omega^{1-\delta}}\,d\omega\\
&+\int_{\omega_{\rm c}}^\infty
\widetilde G(\omega)\frac{1-\cos\omega t}{\omega^{2-\delta}}d\omega.
\end{split}
\end{equation}
Applying the same arguments as for the cases~2 and 3, we conclude that the first and the last terms tend to constants. The second term is analysed analogously to that of the case~2:
\begin{equation*}
\begin{split}
\int_0^{\omega_{\rm c}}
\frac{1-\cos\omega t}{\omega^{2-\delta}}d\omega
&=
\int_0^\infty
\frac{1-\cos\omega t}{\omega^{2-\delta}}d\omega
-\int_{\omega_{\rm c}}^\infty
\frac{1-\cos\omega t}{\omega^{2-\delta}}d\omega
\\&=
t^{1-\delta}\int_0^{\infty}
\frac{1-\cos\upsilon}{\upsilon^{2-\delta}}\,d\upsilon
-\frac1{(1-\delta)\omega_{\rm c}^{1-\delta}}
+\int_{\omega_{\rm c}}^\infty
\frac{\cos\omega t}{\omega^{2-\delta}}d\omega.
\end{split}
\end{equation*}
The last integral converges to zero due to the Riemann-Lebesgue lemma.

Consider the integral in the third term in Eq.~(\ref{EqMainDecompositionG}). It can be calculated explicitly in terms of the generalized hypergeometric functions, but, for our purposes, the following estimations suffice. If $0<\delta<1$, then, using the substitution $\upsilon=\omega t$ again, we can perform as follows:
\begin{equation*}
\int_0^{\omega_{\rm c}}
\frac{1-\cos\omega t}{\omega^{1-\delta}}d\omega
=
\frac{\omega_{\rm c}^\delta}\delta-
t^{-\delta}\int_0^{\omega_{\rm c}t}
\frac{\cos\upsilon}{\upsilon^{1-\delta}}d\upsilon.
\end{equation*}
The last integral converges to a constant (due to Dirichlet's test), while $t^{-\delta}$ tends to zero as $t\to\infty$. If $-1<\delta\leq0$ (actually, the case $\delta=0$ corresponds to case~2, but we include it also here), then we have
\begin{equation*}
\int_0^{\omega_{\rm c}}
\frac{1-\cos\omega t}{\omega^{1-\delta}}d\omega
=
t^{-\delta}\int_0^{\omega_{\rm c}t}
\frac{1-\cos\upsilon}{\upsilon^{1-\delta}}d\upsilon
\leq
t^{-\delta}\int_0^1
\frac{1-\cos\upsilon}{\upsilon^{1-\delta}}d\upsilon
+
t^{-\delta}\int_1^{\omega_{\rm c}t}
\frac{1-\cos\upsilon}{\upsilon}d\upsilon.
\end{equation*}
The first integral here is a constant. The second integral was analysed before: Its principal term is $\ln t$ as $t\to\infty$. Again, the expressions for the constants follow from these calculations and from the following:
\begin{equation*}
\begin{split}
\widetilde G(0)&=
\lim_{\omega\to+0}
G(\omega)\omega\coth\frac{\beta\omega}2
=
2\beta^{-1}G(0),
\\
\widetilde G'(0)&=
\lim_{\omega\to+0}
\left(
G(\omega)\omega\coth\frac{\beta\omega}2
\right)'
=
2\beta^{-1}G'(0).
\end{split}
\end{equation*}
This finishes the prove of the last case~4 of the theorem. 
\end{proof}

\section{Two examples of the Ohmic spectral densities}\label{SecExamples}

Let us consider two popular choices of Ohmic spectral densities. The first choice is the exponential cutoff:
\begin{equation*}
J(\omega)=\omega e^{-\omega/\Omega},
\end{equation*}
where $\Omega$ is the characteristic frequency of the bath.
Here $\Gamma_0=\pi\beta^{-1}$ and $\alpha=-2(\beta\Omega)^{-1}$, hence, 
\begin{equation}\label{EqDecohExample}
\varrho_{10}(t)=\varrho_{10}(0)B(t)
t^{2(\beta\Omega)^{-1}}
e^{-\pi\beta^{-1} t}.
\end{equation}
The second choice is the Drude-Lorentz spectral density:
\begin{equation*}
J(\omega)=\frac{\omega\Omega^2}{\omega^2+\Omega^2}.
\end{equation*}
Here $\Gamma_0=\pi\beta^{-1}$ and $\alpha=0$, hence,
\begin{equation*}
\varrho_{10}(t)=\varrho_{10}(0)B(t)e^{-\pi\beta^{-1} t}.
\end{equation*}
Thus, in the latter case, the decoherence is slightly faster due to different values of $J_{\rm eff}'(0)$. If $\Omega$ is large, then the difference is negligible, but if $\Omega$ is small, then the difference can be significant. 

Generally, large value of $|J_{\rm eff}'(0)|$ (which corresponds to either a sharp peak or a sharp hollow at zero) can significantly modify the rate of decoherence. 

Also it is interesting that the long-term dynamics of coherences is determined by only two values: $\lim_{\omega\to+0}J(\omega)/\omega$ and $\lim_{\omega\to+0}[J(\omega)/\omega]'$ and does not depend on further details of the function $J(\omega)$.

\section{Discussion of the Markovian (weak-coupling) limit}
\label{SecWeak}

In the theory of weak system-bath coupling limit, one considers a Hamiltonian where the interaction Hamiltonian is multiplied by a small dimensionless parameter $\lambda$:
\begin{equation}
H=H_{\rm S}+H_{\rm B}+\lambda H_{\rm I},\quad\lambda\to0.
\end{equation}
One can derive the Davies quantum dynamical semigroup (or, equivalently, the Davies Markovian quantum master equation) for the reduced density operator of the system \cite{Davies,Davies2,MerkliRev}. It predict that, for the considered model, only $J_{\rm eff}(0)$ matters: if this value is positive (Ohmic spectral density), then the exponential full decoherence takes place and the rate of decoherence is proportional to $\lambda^2$. In the super-Ohmic case, $J_{\rm eff}(0)=0$ and neither full nor partial  decoherence occurs. 

Let us analyse this from the point of view of the presented analysis. We should replace $J(\omega)$ and, thus, $\Gamma(t)$ by $\lambda^2 J(\omega)$ and $\lambda^2\Gamma(t)$, respectively. Then, if $J(\omega)\sim c\omega^{2+\delta}$ as $\omega\to+0$ for some $c>0$, the partial decoherence (\ref{EqGammaConst}) and (\ref{EqDecohPartial}) is negligible in this limit. Indeed, $\Gamma(t)$ is a bounded function in this case and $\lambda^2\Gamma(t)$ tends to zero uniformly on $[0,\infty)$, hence, $\varrho_{10}(t)$ uniformly tends to $\varrho_{10}(0)$.

If the spectral density is Ohmic, then the Davies quantum dynamical semigroup correctly predicts the exponent in (\ref{EqDecohExp}). In particular, the decoherence rate is indeed proportional to $\lambda^{2}$. But the Markovian master equation do not capture the power term $t^\alpha$. From the point of view of the formal limit $\lambda\to0$, the influence of the power-law terms disappear in the weak-coupling limit. This can be seen as follows: If $\tau=(\lambda^2\Gamma_0)^{-1}$ is the characteristic time of decay of the exponential term, then $\tau^{\lambda^2\alpha}=(\lambda^2\Gamma_0)^{-\lambda^2\alpha}\to1$ as $\lambda\to0$. So, in this limit, the power-law terms significantly differ from 1 only on the times where the coherence is already suppressed by the exponential term. This formally justifies the Davies quantum dynamical semigroup. However, if we consider a concrete physical system, then $\lambda$ can be small, but is a constant. Then, a sharp peak or hollow at zero (large value of $|J'_{\rm eff}(0)|$) can significantly modify the decoherence rate predicted by the Davies master equation. There is a particular case of a known fact that the Markovian approximation does not work in the case of rapid changes of the spectral density near the Bohr frequencies (differences between the energy levels). Here only the zero Bohr frequency is important.

In the case of the sub-Ohmic spectral density $J(\omega)\sim c\omega^{1+\delta}$ as $\omega\to+0$, where $-1<\delta\leq0$ and $c>0$, the decoherence is superexponential and the Davies quantum master equation cannot describe it since $J_{\rm eff}(\omega)\to\infty$ as $\omega\to+0$ (which reflects a superexponential law of decoherence). 

The ``mild'' super-Ohmic case $J(\omega)\sim c\omega^{1+\delta}$ as $\omega\to+0$ for $0<\delta\leq1$ and $c>0$ requires a bit more attention. In this case, the decoherence is slower than exponential, which also cannot be captured by the Markovian master equation: The full decoherence takes place, but the Markovian master equation predicts no decoherence. 

Namely, let us denote $\overline \varrho_{10}(t)$ the coherence predicted by the Davies master equation (in contrast to the exact value $\varrho_{10}(t)$). In our case, we have $\overline \varrho_{10}(t)=\varrho_{10}(0)$ and $\varrho_{10}(t)\to0$ as $t\to\infty$. Nevertheless, the Davies theorem is true, which says (for this particular simple model) \cite{Davies,Davies2} that
\begin{equation}
\lim_{\lambda\to0}\sup_{0\leq t\leq T/\lambda^2}
|\overline \varrho_{10}(t)-\varrho_{10}(t)|=0
\end{equation}
for an arbitrary finite $T$. Indeed, if one of the asymptotic formulas (\ref{EqGammaLog}) or (\ref{EqGammaSubLin}) is valid, then $\lambda^2\Gamma(t)$ tends to 1 uniformly on $[0,T/\lambda^2]$ for any $T>0$. However, increasing $T$ requires making $\lambda$ smaller and smaller to maintain a constant level of error.

In Ref.~\cite{MerkliRev}, using the resonance theory, it is proved that, under certain additional conditions (in comparison with the Davies theorem), the norm of the difference between the exact reduced density operator of the system and that given by the Davies quantum dynamical semigroup are bounded by $C\lambda^2$ for some $C$ uniformly on the whole time half-line $t\in[0,\infty)$. Obviously, this is not true in our case. This is because the mentioned additional conditions are not satisfied in this model. Namely, one of the conditions (the so called Fermi Golden Rule condition) says exactly that $\lambda^{-2}$ is the only characteristic time scale of dissipative dynamics, which is violated in the considered case. So, this comparison can serve as an example showing that the additional conditions in the mentioned theorem strengthening the Davies theorem, are important (``physical'') and not merely ``technical''.

The inclusion of a term proportional to $\sigma_x$ into the system Hamiltonian (i.e., transitions between the energy levels) restores the time scale $\lambda^{-2}$. In this case, the pure decoherence is accompanied by the exponential decoherence due to quantum transitions, which are described within the Markovian master equations. If the non-exponential pure decoherence is suppressed by the exponential decoherence due to transitions, then the error of the Markovian master equations is not large. 

Note also that quantum master equation for the case where the system-bath interaction is not weak, but an additional term propotional to $\sigma_x$ in the system Hamiltonian can be treated as a small perturbation, was proposed \cite{Fay,TrushUltra}. This is the so called strong-decoherence regime: a strong pure decoherence is accompanied by slow transitions between the energy levels in the system.

\section{Problem of the Markovian embedding}\label{SecMarkEmb}

The case of non-exponential decoherence is interesting for one more problem. A popular way of dealing with non-Markovian (non-semigroup) dynamics of an open quantum system is to embed it into a larger system   whose dynamics is Markovian (semigroup); see Fig.~\ref{fig1}. Since the exact system-bath unitary dynamics is already, obviously, Markovian, an additional condition is usually assumed: The enlarged system should be finite-dimensional or, at least, have a discrete energy spectrum (e.g., a finite-dimensional system plus a finite number of harmonic oscillators), so that it is possible to cut high energy levels (depending on the temperature of the bath) and to consider a finite-dimensional enlarged system. The usual physical interpretation of the extension of the system is that only a part of the bath strongly interacting with the system, while the residual bath interacts with both the system and the separated part of the bath only weakly.

Namely, let ${\rm S}$ be a system with a corresponding finite-dimensional Hilbert space $\mathcal H_{\rm S}$ (in our case, $\mathcal H_{\rm S}=\mathbb C^2$). Its open dynamics is given by a family of completely positive and trace-preserving maps $\{\Phi_t\}_{t\geq0}$, so that $\rho_{\rm S}(t)=\Phi_t\rho_{\rm S}(0)$, where $\rho_{\rm S}(t)$ is the density operator of the system. We will speak about the Markovian embedding if $\rho_{\rm S}(t)=\Tr_{\rm S'}\rho_{\rm SS'}(t)$, where $\rho_{\rm SS'}(t)$ is the density operator of the enlarged system with the (also finite-dimensional) Hilbert space $\mathcal H_{\rm S}\otimes\mathcal H_{\rm S'}$ and $\rho_{\rm SS'}(t)$ satisfies a master equation in the GKSL form:
\begin{equation}
\dot\rho_{\rm SS'}(t)=\mathcal L\rho_{\rm SS'}(t)=
-i[H_{\rm SS'},\rho_{\rm SS'}(t)]
+\sum_{k=1}^K\left(
L_k\rho_{\rm SS'}(t)L_k^\dag
-\frac12
\{L_k^\dag L_k,\rho_{\rm SS'}(t)\}
\right).
\end{equation}
Here $H_{\rm SS'}$ is a self-adjoint operator (a Hamiltonian) in $\mathcal H_{\rm S}\otimes\mathcal H_{\rm S'}$, $L_k$ are linear operators in $\mathcal H_{\rm S}\otimes\mathcal H_{\rm S'}$, $[\cdot,\cdot]$ and $\{\cdot,\cdot\}$ are commutator and anti-commutator, respectively. In other words, $\rho_{\rm SS'}(t)=e^{t\mathcal L}\rho_{\rm SS'}(0)$, where $e^{t\mathcal L}$ is the quantum dynamical semigroup with the generator $\mathcal L$ \cite{BP,AL}.

However, in this case, according to the general theory of systems of ordinary differential equations and also general operator theory, $\rho_{\rm SS'}(t)$ and, thus $\rho_{\rm S}(t)$, is a linear combination of terms of the form $t^{n_j}e^{-l_jt}$, where $-l_j$ are the eigenvalues of $\mathcal L$ and $n_j$ are non-negative integers. Moreover, from the positivity of $e^{t\mathcal L}$, we have $\Re l_j\geq0$ and $n_j=0$ whenever $\Re l_j=0$. 

So, such a Markovian embedding can describe only the exponential relaxation of $\rho_{\rm S}(t)$ to a stationary state (up to the factors $t^{n_j}$). In contrast, as we saw, the model of pure decoherence (\ref{EqH}) allow the super-, sub-exponential and power-law decoherence, like $e^{-At\sqrt t}$, $e^{-A\sqrt t}$ and $t^{-\alpha}$, respectively. Such dynamics cannot be reproduced by a Markovian embedding. Even in the case of an Ohmic spectral density, the Markovian embedding cannot reproduce the power-law factor $t^{-\alpha}$ in Eq.~(\ref{EqDecohExp}) [like in example (\ref{EqDecohExample})] if $\alpha$ is not a negative integer. Thus, in this particular model, the Markovian embedding is not excluded only for a very specific class of spectral densities.

It should be noted that the non-exponential decoherence is not exotic. For example, the decoherence law $e^{-At^2}$ is observed in superconducting qubits as a consequence of flicker noise ($1/f$-noise), where the effective spectral density behaves as $J_{\rm eff}(\omega)\sim c/\omega$ as $\omega\to+0$ \cite{QEngSupercond}. With such a spectral density, integral (\ref{EqGamma}) diverges, but we can consider a regularization: $J=\omega^{\varepsilon}G(\omega)$, where $\varepsilon>0$ is small and $G(0)>0$, so that $J_{\rm eff}\sim 2\beta^{-1}G(0)/\omega^{1-\varepsilon}$ as $\omega\to+0$. Then, according to Theorem~\ref{ThMain} (namely, Eq.~(\ref{EqDecohSuperExp})),
\begin{equation*}
\varrho_{10}(t)=\varrho_{10}(0)
e^{-At^{2-\varepsilon}+2\beta^{-1}G'(0)O(t^{1-\varepsilon}\ln t)},
\quad t\to\infty.
\end{equation*}

In Ref.~\cite{StochPseudomodes}, it is shown that the inclusion of a classical noise to a Markovian embedding can reproduce the non-exponential decoherence. In other words, the inclusion of a classical noise (which can be non-Markovian, e.g., the aforementioned flicker noise) significantly extent the power of the method of Markovian embedding.

Two more comments should be made here:
\begin{itemize}
\item We still can hope to reproduce the dynamics of an open quantum system on a \textit{finite} time interval by a Markovian embedding. But this is already not about a physically meaningful representation of the model (the aforementioned separation of the strongly interacting part of the bath), but rather about merely mathematical approximation of the time dependence.

\item A Markovian embedding can be used not only for the approximation of the dynamics, but also for the approximation of the equilibrium state \cite{Segal}, which is non-Gibbsian if the system-bath coupling is not negligibly weak \cite{TMCA}. Here we do not consider this purpose of Markovian embeddings and write only about the problem of approximation of dynamics.
\end{itemize}

\section{Asymptotic Markovianity}\label{SecAsympMark}

In the previous section, we have shown that the Markovian embedding (without additional means like the classical noise) is impossible if the spectral density is not Ohmic and, thus, the decoherence is non-exponential. Let us consider now the case of the Ohmic spectral density and the exponential decoherence, i.e., case~2 of Theorem~\ref{ThMain}. Here, vice versa, the asymptotic Markovianity in the following sense takes place.

The exact solution $\varrho(t)$ (\ref{EqVarrho}) with the function $\Gamma(t)$ (\ref{EqGamma}) obviously satisfies the following equation with a time-dependent GKSL generator:
\begin{equation*}
\dot\varrho=
\frac{\gamma(t)}2(\sigma_z\varrho\sigma_z-\varrho),
\end{equation*}
or, if we return to the Schr\"{o}dinger picture,
\begin{equation*}
\dot\rho_{\rm S}=-i[H_{\rm S},\rho_{\rm S}]
+\frac{\gamma(t)}2(\sigma_z\rho_{\rm S}\sigma_z-\rho_{\rm S}),
\end{equation*}
where $\gamma(t)=\dot\Gamma(t)$. In the case $J(\omega)\sim c\omega$ as $\omega\to+0$, doing in the same manner as in Eq.~(\ref{EqMainDecomposition}), we arrive at
\begin{equation*}
\begin{split}
\gamma(t)&=\dot\Gamma(t)
=\int_0^\infty 
J_{\rm eff}(\omega)\frac{\sin\omega t}\omega\,d\omega
\\
&=\int_0^{\omega_{\rm c}}
\frac{J_{\rm eff}(\omega)-J_{\rm eff}(0)}{\omega}
\sin\omega t\,d\omega
+J_{\rm eff}(0)\int_0^{\omega_{\rm c}}
\frac{\sin\omega t}{\omega}\,d\omega
+\int_{\omega_{\rm c}}^\infty
J_{\rm eff}(\omega)\frac{\sin\omega t}\omega\,d\omega.
\end{split}
\end{equation*}
Again, applying the Riemann-Lebesgue lemma, we obtain $\gamma(t)\to\Gamma_0={\rm const}$ [see Eq.~(\ref{EqGamma0})] as $t\to\infty$ and, thus, for large times, we obtain a GKSL generator of a quantum dynamical semigroup. Of course, this can be seen also directly from Eqs.~(\ref{EqGammaLin}) and (\ref{EqDecohExp}) since the linear term in $\Gamma(t)$ is principal for large times.

This asymptotic Markovianity is an interesting phenomenon. Usually, we expect the semigroup dynamics in the weak-coupling limit (or another limit). However, here we have obtained that, for large times, the semigroup property is satisfied for any system-bath coupling strength. A peculiarity of the weak-coupling limit here is that the transient non-Markovian stage of the dynamics is infinitesimal and can be neglected in the principal order approximation, i.e., the second-order in $\lambda$. We saw it in Sec.~\ref{SecWeak}. Note that, in Ref.~\cite{TereJPA}, it is shown that the effects of the transient dynamics should be included in the higher-order approximations.

However, as we mentioned in the introduction, the semigroup property is not the most general definition of quantum Markovianity. The definition generalizing the corresponding definition from the classical theory of random processes, is based on the quantum regression formula \cite{MarkHier,VacchiniQRT,ChrusBeyondMark,Lax,LoGullo}.

Let $H$, as before, be a system-bath Hamiltonian and the initial system-bath state is the product state $\rho_{\rm S}\otimes\rho_{{\rm B},\beta}$, where $\rho_{\rm S}$ is arbitrary and $\rho_{{\rm B},\beta}$ as the initial state of the bath [see Eq.~(\ref{EqRhoIni})]. One says that the open quantum system satisfies the quantum regression formula if, for every $\rho_{\rm S}$, every $n$ and every bounded operators $X_1,\ldots,X_n,Y_1,\ldots,Y_n$ in the Hilbert space of the system and all $t_1,\ldots, t_n\geq0$, the following equality holds:
\begin{equation}\label{EqQRF}
\Tr_{\rm SB}
\big[
\tilde{\mathcal E}_n\mathcal U_{t_n}
\ldots
\tilde{\mathcal E}_1\mathcal U_{t_1}
(\rho_{\rm S}\otimes\rho_{{\rm B},\beta})
\big]
=
\Tr_{\rm S}
\big[
\mathcal E_n\Phi_{t_n}
\ldots
\mathcal E_1\Phi_{t_1}
(\rho_{\rm S})
\big],
\end{equation}
where $\mathcal U_t=e^{-itH}(\cdot)e^{itH}$, 
$\Phi_t=\Tr_{\rm B}
[\mathcal U_t(\,\cdot\,\otimes\rho_{{\rm B},\beta})]$, and
\begin{equation*}
\tilde{\mathcal E}_k=(X_k\otimes I_{\rm B})(\cdot)(Y_k\otimes I_{\rm B}),
\qquad
\mathcal E_k=X_k(\cdot)Y_k
\end{equation*}
with $I_{\rm B}$ being the identity operator in the bath Hilbert space. The quantum regression formula says that we can use the family of quantum dynamical maps $\{\Phi_t\}_{t\geq0}$ not only for the description of the dynamics of reduced density operator of the system and prediction of the average values of observables, but also for multitime correlation functions. Obviously, the quantum regression formula cannot be satisfied exactly, except trivial cases (e.g., $H_{\rm I}=0$). We can hope for the satisfaction of this formula only in certain limiting cases. In particular, the quantum regression formula is satisfied in the weak-coupling limit \cite{Dumke1983}.

In Ref.~\cite[Proposition~2.2]{LonigroChrus}, it is shown that, for the considered model of pure decoherence, the quantum regression formula (\ref{EqQRF}) is satisfied if and only if the following formula holds for all $n$, $j_1,\ldots,j_n,l_1,\ldots,l_n\in\{0,1\}$, $t_1,\ldots,t_n\geq0$:
\begin{equation}\label{EqQRF2}
\Tr\big[
e^{-it_nH_{j_n}}\ldots e^{-it_1H_{j_1}}
\rho_{{\rm B},\beta}
e^{it_1H_{l_1}}\ldots e^{-it_nH_{l_n}}
\big]
=
\prod_{k=1}^n
\Tr\big[
e^{-it_kH_{j_k}}
\rho_{{\rm B},\beta}
e^{it_kH_{j_k}}
\big],
\end{equation}
where $H_j$ for $j=0,1$ are given in Eq.~(\ref{EqH10}).

\begin{Theorem}\label{ThQRF}
Consider the thermodynamical limit $N\to\infty$ in Eq.~(\ref{EqJ}) such that $J(\omega)$ converges to a function [also denoted by $J(\omega)$] that is integrable on $[0,\infty)$ and twice differentiable on $[0,\omega_1)$ for some $\omega_1>0$ and such that $J(\omega)\sim c\omega$ as $\omega\to+0$ for some $c>0$. Then the quantum regression formula is  asymptotically satisfied for large times in the following sense:

\begin{equation}\label{EqQRFleft}
\lim_{N\to\infty}
\Tr\big[
e^{-it_nH_{j_n}}\ldots e^{-it_1H_{j_1}}
\rho_{{\rm B},\beta}
e^{it_1H_{l_1}}\ldots e^{-it_nH_{l_n}}
\big]
=e^{-\Gamma_0 T+O(\ln(\min t))},
\quad \min t\to\infty,
\end{equation}
and
\begin{equation}\label{EqQRFright}
\lim_{N\to\infty}
\prod_{k=1}^n
\Tr\big[
e^{-it_kH_{j_k}}
\rho_{{\rm B},\beta}
e^{it_kH_{l_k}}
\big]
=e^{-\Gamma_0 T+O(\ln(\min t))},
\quad \min t\to\infty.
\end{equation}
Here 
\begin{equation*}
T=\sum_{k\colon j_k\neq l_k}t_k, 
\qquad
\min t=\min_{k\colon j_k\neq l_k}t_k,
\end{equation*}
and $\Gamma_0$ is given in Eq.~(\ref{EqGamma0}).
Thus, both sides of Eq.~(\ref{EqQRF2}) are exponential functions whose arguments coincide in the principal terms as the minimal $t_k$ goes to infinity.
\end{Theorem}
Note that the first limit in Eqs.~(\ref{EqQRFleft}) and (\ref{EqQRFright}) is the thermodynamic, i.e., $N\to\infty$, and the second one is $\min t\to\infty$.

\begin{proof}[Proof of Theorem~\ref{ThQRF}]
Let us calculate (\ref{EqQRFright}) first. If $j_k=l_k$, then 
$\Tr\big[
e^{-it_kH_{j_k}}
\rho_{{\rm B},\beta}
e^{it_kH_{l_k}}
\big]
=1$. If $j_k\neq l_k$, then, according to Eqs.~(\ref{EqExpGamma}) and (\ref{EqGammaLin}),
$$\Tr\big[
e^{-it_kH_{j_k}}
\rho_{{\rm B},\beta}
e^{it_kH_{l_k}}
\big]
=e^{-\Gamma_0 t_k+O(\ln t_k)},\quad t_k\to\infty.$$ 
Taking the product over all $k$, we obtain Eq.~(\ref{EqQRFright}).

For the proof of Eq.~(\ref{EqQRFleft}), we will use the results of calculations from Ref.~\cite[Proof of Proposition~3.2]{LonigroChrus}. According to them, the left-hand side of Eq.~(\ref{EqQRFleft}) is equal to
\begin{equation*}
\exp\left\lbrace
-\frac12
\int_0^\infty \frac{J_{\rm eff}(\omega)}{\omega^2}
\left|
\sum_{k=1}^n(j_k-l_k)(e^{i\omega T_k}-e^{i\omega T_{k-1}})
\right|^2
d\omega
\right\rbrace,
\end{equation*}
where $T_k=\sum_{k'=1}^k t_{k'}$. Now we apply decomposition (\ref{EqMainDecomposition}):
\begin{equation}\label{EqMainDecompositionQRF}
\begin{split}
&\int_0^\infty \frac{J_{\rm eff}(\omega)}{\omega^2}
\left|
\sum_{k=1}^n(j_k-l_k)(e^{i\omega T_k}-e^{i\omega T_{k-1}})
\right|^2
d\omega
\\
&=\int_0^{\omega_{\rm c}}
\frac{J_{\rm eff}(\omega)-J_{\rm eff}(0)-J'_{\rm eff}(0)\omega}{\omega^2}
\left|
\sum_{k=1}^n(j_k-l_k)(e^{i\omega T_k}-e^{i\omega T_{k-1}})
\right|^2
d\omega\\
&+J_{\rm eff}(0)\int_0^\infty
\frac1{\omega^2}
\left|
\sum_{k=1}^n(j_k-l_k)(e^{i\omega T_k}-e^{i\omega T_{k-1}})
\right|^2
d\omega\\
&-J_{\rm eff}(0)\int_{\omega_{\rm c}}^\infty
\frac1{\omega^2}
\left|
\sum_{k=1}^n(j_k-l_k)(e^{i\omega T_k}-e^{i\omega T_{k-1}})
\right|^2
d\omega\\
&+J'_{\rm eff}(0)\int_0^{\omega_{\rm c}}
\frac{1}\omega
\left|
\sum_{k=1}^n(j_k-l_k)(e^{i\omega T_k}-e^{i\omega T_{k-1}})
\right|^2
d\omega\\
&+\int_{\omega_{\rm c}}^\infty
\frac{J_{\rm eff}(\omega)}{\omega^2}
\left|
\sum_{k=1}^n(j_k-l_k)(e^{i\omega T_k}-e^{i\omega T_{k-1}})
\right|^2\
d\omega
\end{split}
\end{equation}
for some $\omega_{\rm c}>0$. Here, for all terms in the right-hand side, except the second one, we can apply the following estimation:
\begin{equation*}
\begin{split}
\left|
\sum_{k=1}^n(j_k-l_k)(e^{i\omega T_k}-e^{i\omega T_{k-1}})
\right|^2
&\leq
\sum_{k=1}^n
|j_k-l_k|\cdot
|e^{i\omega T_k}-e^{i\omega T_{k-1}}|^2
\\
&\leq
2\sum_{k=1}^n
|j_k-l_k|\cdot
(1-\cos\omega t_k).
\end{split}
\end{equation*}
Thus, we can repeat the corresponding steps of the proof of the case~2 of Theorem~\ref{ThMain} and to conclude that all these terms are $O(\ln(\min t))$ as $\min t\to\infty$. The integral in the second term in the right-hand side of Eq.~(\ref{EqMainDecompositionQRF}) was shown in Ref.~\cite{LonigroChrus} to be equal to $\pi T$, which concludes the proof of Eq.~(\ref{EqQRFleft}) and, thus, the theorem.
\end{proof}

In Ref.~\cite{LonigroChrus}, D.~Lonigro and D.~Chru\'{s}ci\'{n}ski show that the dynamics is exactly Markovian if we formally replace the lower limit of integration in Eq.~(\ref{EqGamma}) for $\Gamma(t)$ by $-\infty$ and put $J_{\rm eff}(\omega)$ to be constant on the whole line $\omega\in(-\infty,\infty)$. Of course, negative frequencies are unphysical. However, the authors write:
``We point out that, while these choices of coupling may be considered unphysical, the corresponding
results are indicative of what would be obtained in more realistic scenarios: we can
expect an exponential dephasing in the regime in which the spin-boson interaction is `approximately flat' in the energy regime of interest.'' Here we actually show that the spectral density can be considered to be `approximately flat' on large times, when only the behaviour of $J_{\rm eff}(\omega)$ in a vicinity of zero matters.

\section{The case of zero temperature}\label{SecZero}

We have analysed a realistic case of a positive temperature. However, it is worthwhile to briefly mention the case of the zero temperature as well since it is often used as an approximation for the case of low temperatures. In this case, we still have formula (\ref{EqExpGamma}) for decoherence, but $\Gamma(t)$ is defined without the factor $\coth(\beta\omega/2)$ in the integral \cite{BP,AlickiDecoh,EkertQCompDiss,LonigroChrus}:

\begin{equation}\label{EqGammaZero}
\Gamma(t)=
\int_0^\infty J(\omega)
\frac{1-\cos\omega t}{\omega^2}d\omega.
\end{equation}

\begin{Theorem}\label{ThMainZero}

Let $J(\omega)$ be an integrable function on $[0,\infty)$. 

\begin{enumerate}
\item If $J(\omega)\sim c\omega^{1+\delta}$ as $\omega\to+0$, where $\delta>0$ and $c>0$, then 
\begin{equation}\label{EqGammaConstZero}
\lim_{t\to\infty}\Gamma(t)=
\int_0^\infty \frac{J(\omega)}{\omega^2}\coth\left(\frac{\beta\omega}2\right)d\omega\equiv\Gamma_{\infty},
\end{equation}
hence, the decoherence is partial:
\begin{equation}\label{EqDecohPartialZero}
\varrho_{10}(t)=\varrho_{10}(0)B(t)e^{-\Gamma_\infty},
\end{equation}
where $B(t)$ is a bounded function converging to 1 as $t\to\infty$.

\item If $J(0)=c>0$ and there exits $J'(0)$, then
\begin{equation}\label{EqGammaLinZero}
\Gamma(t)=\Gamma_0t+\alpha\ln t+C+o(1),\quad t\to\infty,
\end{equation}
where $C$ is a constant, hence, the decoherence is exponential:
\begin{equation}\label{EqDecohExpZero}
\varrho_{10}(t)=\varrho_{10}(0)B(t) e^{-\Gamma_0 t}/t^\alpha.
\end{equation}
Here
\begin{equation}\label{EqGamma0Zero}
\Gamma_0=\frac\pi2 J(0),\qquad
\alpha=J'(0)
\end{equation}
and $B(t)$ is a bounded function converging to $e^{-C}$ as $t\to\infty$.

\item If $J(\omega)\sim c\omega$ as $\omega\to+0$, where $c>0$, then
\begin{equation}\label{EqGammaLogZero}
\Gamma(t)=\alpha\ln t+C+o(1),\quad t\to\infty,
\end{equation}
hence, the decoherence obeys a power law: 
\begin{equation}\label{EqDecohPowerZero}
\varrho_{10}(t)=\varrho_{10}(0)B(t)/t^\alpha,
\end{equation}
where, again, $B(t)$ is a bounded function converging to $e^{-C}$ as $t\to\infty$.

\item If $J(\omega)=\omega^{\gamma}G(\omega)$ as $\omega\to+0$, where $-1<\gamma<1$, $G(0)>0$, and there exits $G'(0)$, then
\begin{equation}\label{EqGammaSubLinZero}
\Gamma(t)=At^{1-\gamma}+\widetilde C+o(1),\quad t\to\infty,
\end{equation}
if $0<\gamma<1$, and
\begin{equation}\label{EqGammaSuperLinZero}
\Gamma(t)=
At^{1-\gamma}+G'(0)
O(t^{-\gamma}\ln t),\quad t\to\infty,
\end{equation}
if $-1<\gamma\leq0$,
where
\begin{equation*}
A=
G(0)
\int_0^{\infty}
\frac{1-\cos\upsilon}{\upsilon^{2-\gamma}}\,d\upsilon
\end{equation*}
and $\widetilde C$ is a constant.
Thus, the decoherence is subexponential for $\delta>0$ and superexponential for $\gamma<0$:
\begin{equation}\label{EqDecohSubExpZero}
\varrho_{10}(t)=\varrho_{10}(0)\widetilde B(t)e^{-At^{1-\gamma}}
\end{equation}
if $0<\gamma<1$, where $\widetilde B(t)$ is a bounded function converging to $e^{-\widetilde C}$ as $t\to\infty$, and
\begin{equation}\label{EqDecohSuperExpZero}
\varrho_{10}(t)=\varrho_{10}(0)
e^{-At^{1-\gamma}+G'(0) O(t^{-\delta}\ln t)},
\quad t\to\infty
\end{equation}
if $-1<\delta\leq0$.
\end{enumerate}
\end{Theorem}

The results of Theorem~\ref{ThMainZero} can be summarized as follows. Let $J(\omega)\sim c\omega^\gamma$ as $\omega\to+0$ for some $c>0$.
\begin{itemize}
\item If $-1<\gamma<0$ (sub-Ohmic spectral density), then the decoherence is full and its rate is superexponential.

\item If $\gamma=0$ (sub-Ohmic spectral density), then the decoherence is full and its rate is exponential.

\item If $0<\gamma<1$ (sub-Ohmic spectral density), then the decoherence is full and its rate is subexponential, but faster than any degree of $t$.

\item If $\gamma=1$ (Ohmic spectral density), then the decoherence is full and obeys a power law.

\item If $\gamma>1$ (super-Ohmic spectral density), then the decoherence is partial. 
\end{itemize}

The proof of Theorem~\ref{ThMainZero} is completely analogous to that of Theorem~\ref{ThMain}: we simply replace $J_{\rm eff}=J(\omega)\coth(\beta\omega/2)$ by $J(\omega)$ everywhere. In other words, $J_{\rm eff}(\omega)= J(\omega)$ for $\beta=\infty$. Since $\coth(\beta\omega/2)\sim(\beta\omega/2)^{-1}$ as $\omega\to+0$, elimination of this factor leads to the reduction of $\gamma$ by 1 in all cases. For example, the case $J_{\rm eff}(0)=c>0$ corresponds to the exponential decoherence for both positive temperature and zero temperature cases. However, in the case of a positive temperature, this corresponds to $J(\omega)\sim (\beta/2)c\omega$ as $\omega\to+0$, while, in the case of the zero temperature, this corresponds to $J(\omega)=c$. 

Again, if the decoherence is not exponential ($\gamma\neq0$), then a Markovian embedding is impossible, while, in the exponential case $\gamma=0$, the asymptotic Markovianity takes place.

Note that exact form of $\Gamma(t)$ for the particular super-Ohmic spectral density
\begin{equation}\label{EqExampleZero}
J(\omega)=2\eta \omega_{\rm c}^{1-\gamma}\omega^\gamma e^{-\omega/\omega_{\rm c}}
\end{equation}
for $\gamma>1$, where $\eta, \omega_{\rm c}>0$, and the zero temperature was obtained in Ref.~\cite{SuperOhmic}. From this formula, the partial decoherence follows (which is discussed in the mentioned paper), in agreement with Theorem~\ref{ThMainZero}.

Let us consider now the case of a positive, but a very small temperature. How the predictions of Theorems~\ref{ThMain} and~\ref{ThMainZero} can agree? Consider, for example, the aforementioned spectral density (\ref{EqExampleZero}) for $1<\gamma\leq2$. Theorem~\ref{ThMainZero} predicts only a partial decoherence, while Theorem~\ref{ThMain} predicts the full decoherence (though slower than exponential) for this case. Of course, the answer is in different time scales: all decoherence constants in Theorem~\ref{ThMain} are proportional to the temperature $\beta^{-1}$, while those in Theorem~\ref{ThMainZero} are not. So, at the beginning, the system partially decohere and then, on a larger time scale proportional to $\beta$, the full decoherence occurs. Different time scales caused by vacuum and positive-temperature contributions are well-known for this model \cite{BP,EkertQCompDiss,ViolaLloyd1998}.

\section{Conclusions}

The main result of this paper is Theorem~\ref{ThMain} (for the case of a positive temperature of the bath) and also Theorem~\ref{ThMainZero} (for the case of the zero temperature), where the long-time rate of decoherence in a known exactly solvable model of decoherence have been related to the asymptotic behaviour of the bath spectral density at low frequencies. Though the considered model of pure decoherence is paradigmatic in the theory of open quantum systems, we are not aware of a such detailed analysis of the long-term behaviour of coherence in this model.

We have discussed consequences of these results for the theory of weak-coupling limit, for the possibility of the Markovian embedding, and for the asymptotic Markovianity. In particular, we see that the decoherence is not necessarily exponential, while Markovian embeddings can give only the exponential relaxation to a steady state. Hence, the Markovian embedding, which is widely used to model the non-Markovian dynamics of open quantum systems, is not always possible. From the other side, if the spectral density is Ohmic (and the temperature is positive), the exponential decoherence dominates for large times. As a consequence, we have asymptotic Markovianity in the most general sense: in the sense of the quantum regression formula.

Finally, it is worthwhile to mention works about theoretical analysis of decoherence in the considered model with a special control technique called the dynamical decoupling, which is aimed to suppress the decoherence \cite{ViolaLloyd1998,Viola2013}. It would be interesting to extend the results of the present paper about the long-term rates of decoherence depending on the asymptotic behaviour of the spectral density at low frequencies to the case of dynamical decoupling.

\vspace{6pt}

\funding{This work was funded by the Russian Federation represented by the Ministry of Science and Higher Education of the Russian Federation (grant number 075-15-2020-788).}


\acknowledgments{The author thanks Alexander Teretenkov, Dvira Segal, and Michiel Burgelman for fruitful discussions on the subject of the present paper and its results.}

\conflictsofinterest{The author declares no conflict of interest.}

\abbreviations{Abbreviations}{
The following abbreviation is used in this manuscript:\\

\noindent 
\begin{tabular}{@{}ll}
GKSL & Gorini-Kossakowski-Sudarshan-Lindblad
\end{tabular}
}


\reftitle{References}


%



\begin{thebibliography}{999}

\bibitem{BP}
Breuer, H.-P.; Petruccione, F. \textit{The Theory of Open Quantum Systems}; Oxford University Press: New York, NY, USA, 2002.

\bibitem{EkertQCompDiss}
Palma, G.M.; Suominen K.-A.; Ekert, A.K.
Quantum computers and dissipation,
{\em Proc. Roy. Soc. Lond. A} {\bf 1996}, {\em 452}, 567--584. 
[\href{https://doi.org/10.1098/rspa.1996.0029}{CrossRef}]

\bibitem{AlickiDecoh}
Alicki, R.
Pure decoherence in quantum systems,
{\em Open Syst. Inf. Dyn.} {\bf 2004}, {\em 11}, 53--61. 
[\href{https://doi.org/10.1023/B:OPSY.0000024755.58888.ac}{CrossRef}]

\bibitem{Brito}
Brito, F.; Werlang, T.
A knob for Markovianity,
{\em New J. Phys.} {\bf 2015}, {\em 17}, 072001. 
[\href{https://doi.org/10.1088/1367-2630/17/7/072001}{CrossRef}]

\bibitem{VacchiniQRT}
Guarnieri, G.; Smirne, A.; Vacchini, B.
Quantum regression theorem and non-Markovianity of quantum dynamics,
{\em Phys. Rev. A} {\bf 2014}, {\em 90}, 022110. 
[\href{https://doi.org/10.1103/PhysRevA.90.022110}{CrossRef}]


\bibitem{MerkliNesterovDimer}
Merkli, M.; Berman, G.P.; Sayre, R.T.; Gnanakaran, S.; 
K\"{o}nenberg, M.; Nesterov, A.\,I.; Song, H.
Dynamics of a chlorophyll dimer in collective and local thermal environments,
{\em J. Math. Chem.} {\bf 2016}, {\em 54}, 866--917. 
[\href{https://doi.org/10.1007/s10910-016-0593-z}{CrossRef}]


\bibitem{LonigroChrus}
Lonigro, D.; Chru\'{s}ci\'{n}ski, D. 
Quantum regression in dephasing phenomena, 
{\em J. Phys. A} {\bf 2022}, {\em 55}, 225308. 
[\href{https://doi.org/10.1088/1751-8121/ac6a2d}{CrossRef}]

\bibitem{SuperOhmic}
Nacke, Ph.; Otterpohl, F.; Thorwart, M.; Nalbach P.
Quantum regression theorem and non-Markovianity of quantum dynamics,
{\em Phys. Rev. A} {\bf 2023}, {\em 107}, 062218. 
[\href{https://doi.org/10.1103/PhysRevA.107.062218}{CrossRef}]

\bibitem{MarkHier}
Li, L.; Hall, M.J.W.; Wiseman, H.M. 
Concepts of quantum non-Markovianity: a hierarchy,
{\em Phys. Rep.} {\bf 2018}, {\em 759}, 1--51. 
[\href{https://doi.org/10.1016/j.physrep.2018.07.001}{CrossRef}]


\bibitem{ChrusIntroNonMark}
Chru\'{s}ci\'{n}ski, D. 
Introduction to non-Markovian evolution of $n$-level quantum systems.
In {\em Open Quantum Systems. A Mathematical Perspective}; Bahns, D.; Pohl, A.; Witt, I., Eds.; Springer Nature Switzerland, 2010; pp. 55--76.


\bibitem{ChrusBeyondMark}
Chru\'{s}ci\'{n}ski, D. 
Dynamical maps beyond Markovian regime 
Available online: URL 
\href{https://arxiv.org/abs/2209.14902}
{https://arxiv.org/abs/2209.14902} 
(accessed on 07th November 2023).


\bibitem{Lax}
Lax, M.
Formal theory of quantum fluctuations from a driven state,
{\em Phys. Rev.} {\bf 1963}, {\em 129}, 2342--2348. 
[\href{https://doi.org/10.1103/PhysRev.129.2342}{CrossRef}]

\bibitem{LoGullo}
Lo Gullo, N.; Sinayskiy, I.; Busch, Th.; Petruccione, F.
Non-Markovianity criteria for open system dynamics 
Available online: URL 
\href{https://arxiv.org/abs/1401.1126}
{https://arxiv.org/abs/1401.1126} 
(accessed on 07th November 2023).

\bibitem{Tamapre}
Tamascelli, D.; Smirne, A.; Lim, J.; Huelga, S.F.; Plenio M.B.
Efficient simulation of finite-temperature open quantum systems,
{\em Phys. Rev. Lett.} {\bf 2019}, {\em 123}, 090402. 
[\href{https://doi.org/10.1103/PhysRevLett.123.090402}{CrossRef}]

\bibitem{Tama}
Mascherpa, F.; Smirne, A.; Somoza, A.D.; Fern\'{a}ndez-Acebal, P.; Donadi, S.; Tamascelli, D.; Huelga, S.F.; Plenio, M.B.
Optimized auxiliary oscillators for the simulation of general open quantum systems,
{\em Phys. Rev. A} {\bf 2020}, {\em 101}, 052108. 
[\href{https://doi.org/10.1103/PhysRevA.101.052108}{CrossRef}]

\bibitem{GarrawayPetruc}
Pleasance, G.; Garraway, B.M.; Petruccione, F.
Generalized theory of pseudomodes for exact descriptions of non-Markovian quantum processes,
{\em Phys. Rev. Res.} {\bf 2020}, {\em 2}, 043058. 
[\href{https://doi.org/10.1103/PhysRevResearch.2.043058}{CrossRef}]

\bibitem{TereFinT}
Teretenkov, A.E.
Integral representation of finite temperature non-Markovian evolution of some systems in rotating wave approximation,
{\em Lobachevskii J. Math.} {\bf 2020}, {\em 41}, 2397--2404. 
[\href{https://doi.org/10.1134/S1995080220120410}{CrossRef}]

\bibitem{TereSeveralBath}
Teretenkov, A.E.
Exact non-Markovian evolution with several reservoirs,
{\em Phys. Part. Nucl.} {\bf 2020}, {\em 51}, 479--484. 
[\href{https://doi.org/10.1134/S1063779620040711}{CrossRef}]


\bibitem{Lambert}
Iles-Smith, J.; Lambert, N.; Nazir, A.
Environmental dynamics, correlations, and the emergence of noncanonical equilibrium states in open quantum systems,
{\em Phys. Rev. A} {\bf 2014}, {\em 90}, 032114. 
[\href{https://doi.org/10.1103/PhysRevA.90.032114}{CrossRef}]


\bibitem{Strasberg}
Strasberg, P.; Schaller, G.; Lambert, N.; Brandes, T.
Nonequilibrium thermodynamics in the strong coupling and non-Markovian regime based on a reaction coordinate mapping,
{\em New J. Phys.} {\bf 2016}, {\em 18}, 073007. 
[\href{https://doi.org/10.1088/1367-2630/18/7/073007}{CrossRef}]

\bibitem{Segal}
Anto-Sztrikacs, N.; Nazir, A.; Segal, D.
Effective Hamiltonian theory of open quantum systems at strong coupling,
{\em PRX Quantum} {\bf 2023}, {\em 4}, 020307. 
[\href{https://doi.org/10.1103/PRXQuantum.4.020307}{CrossRef}]

\bibitem{Luchnikov19}
Luchnikov, I.A.; Vintskevich, S.V.; Ouerdane, H.; Filippov, S.N.
Simulation complexity of open quantum dynamics: Connection with tensor networks,
{\em Phys. Rev. Lett.} {\bf 2019}, {\em 122}, 160401. 
[\href{https://doi.org/10.1103/PhysRevLett.122.160401}{CrossRef}]

\bibitem{Luchnikov22}
Luchnikov, I.A.; Kiktenko, E.O.; Gavreev, M.A.; Ouerdane, H.; Filippov, S.N.; Fedorov A.K.
Probing non-Markovian quantum dynamics with data-driven analysis: Beyond ``black-box'' machine-learning models,
{\em Phys. Rev. Research} {\bf 2022}, {\em 4}, 043002. 
[\href{https://doi.org/10.1103/PhysRevResearch.4.043002}{CrossRef}]


\bibitem{Davies}
Davies, E.B.
Markovian master equations,
{\em Commun. Math. Phys.} {\bf 1974}, {\em 39}, 91--110. 
[\href{https://doi.org/10.1007/BF01608389}{CrossRef}]

\bibitem{Davies2}
Davies, E.B.
Markovian master equations. II,
{\em Math. Ann.} {\bf 1976}, {\em 219}, 147--158. 
[\href{https://doi.org/10.1007/BF01351898}{CrossRef}]

\bibitem{MerkliRev}
Merkli, M.
Quantum Markovian master equations: Resonance theory shows validity for all time scales,
{\em Ann. Phys.} {\bf 2020}, {\em 412}, 167996. 
[\href{https://doi.org/10.1016/j.aop.2019.167996}{CrossRef}]


\bibitem{TMCA}
Trushechkin, A.S.; Merkli, M.; Cresser, J.D.; Anders, J. 
Open quantum system dynamics and the mean force Gibbs state. 
{\em AVS Quantum Sci.} {\bf 2022}, {\em 4}, 012301. 
[\href{https://doi.org/10.1116/5.0073853}{CrossRef}]

\bibitem{MerkliIdealQGas}
Merkli, M. The ideal quantum gas. In {\em Open Quantum Systems I. The Hamiltonian Approach}; Attal, S.; Joye, A.; Pillet, C.-A., Eds.; Springer: Berlin, Germany, 2006; pp. 183--233.

\bibitem{Viola2013}
Khodjasteh, K.; Sastrawan, J.; Hayes, D.; Green, T.J.; 
Biercuk, M.J.; Viola, L.
Designing a practical high-fidelity long-time quantum memory,
{\em Nature Comm.} {\bf 2013}, {\em 4}, 2045. 
[\href{https://doi.org/10.1038/ncomms3045}{CrossRef}]



\bibitem{GR}
Gradshteyn, I.S.; Ryzhik, I.M.
\textit{Table of Integrals, Series, and Products};
Elsevier: Burlinglon, MA, USA, 2007.


\bibitem{Fay}
Fay, T.P.; Lindoy, L.P.; Manolopoulos, D.E. 
Spin-selective electron transfer reactions of radical pairs: Beyond the Haberkorn master equation,
{\em J. Chem. Phys.} {\bf 2018}, {\em 149}, 064107. 
[\href{https://doi.org/10.1063/1.5041520}{CrossRef}]

\bibitem{TrushUltra}
Trushechkin, A. 
Quantum master equations and steady states for the ultrastrong-coupling limit and the strong-decoherence limit, 
{\em Phys. Rev. A} {\bf 2022}, {\em 106}, 042209. 
[\href{https://doi.org/10.1103/PhysRevA.106.042209}{CrossRef}]
 
\bibitem{AL}
Alicki, R.; Lendi, K. 
{\it Quantum Dynamical Semigroups and Applications}; 
Springer: Berlin, 2007.


\bibitem{QEngSupercond}
Krantz, P.; Kjaergaard, M.; Yan, F.; Orlando, T.P.; Gustavsson S.;  
Oliver, W.D. 
A Quantum Engineer's Guide to Superconducting Qubits, 
{\em Appl. Phys. Rev.} {\bf 2019}, {\em 6}, 021318. 
[\href{https://doi.org/10.1063/1.5089550}{CrossRef}]

\bibitem{StochPseudomodes}
Luo, S.; Lambert, N.; Liang, P.; Cirio M.
Quantum-classical decomposition of Gaussian quantum environments: A stochastic pseudomode model,
{\em PRX Quantum} {\bf 2023}, {\em 4}, 030316. 
[\href{https://doi.org/10.1103/PRXQuantum.4.030316}{CrossRef}]

\bibitem{TereJPA}
Teretenkov, A.E. Non-perturbative effects in corrections to quantum master equation arising in Bogolubov--van Hove limit,
{\em J. Phys. A} {\bf 2021}, {\em 54}, 265302. 
[\href{https://doi.org/10.1088/1751-8121/ac0201}{CrossRef}]

\bibitem{Dumke1983}
D\"{u}mke, R. 
Convergence of multitime correlation functions in the weak and singular coupling limits,
{\em J. Math. Phys.} {\bf 1983}, {\em 24}, 311--315. 
[\href{https://doi.org/10.1063/1.525681}{CrossRef}]

\bibitem{ViolaLloyd1998}
Viola, L.; Lloyd, S. 
Dynamical suppression of decoherence in two-state quantum systems, 
{\em Phys. Rev. A} {\bf 1998}, {\em 58}, 2733--2744. 
[\href{https://doi.org/10.1103/PhysRevA.58.2733}{CrossRef}]



\end{thebibliography}
\end{document}